\pdfoutput=1
\documentclass[aps,twocolumn,superscriptaddress,groupedaddress,showkeys,showpacs]{revtex4-2}
\usepackage{amsmath, amssymb}
\usepackage{amsthm}
\usepackage{graphicx}
\usepackage{physics}
\usepackage{xcolor}
\usepackage{caption}
\usepackage{tikz}
\usetikzlibrary{quantikz2}
\usepackage{float}
\usepackage{amsfonts}
\usepackage{verbatim}
\usepackage{hyperref}
\usepackage[capitalise, nameinlink]{cleveref}

\newcommand*\yl[1]{\textcolor{orange}{\textbf{YL: #1}}}
\newcommand*\jb[1]{\textcolor{blue}{\textbf{JB: #1}}}
\newcommand*\gz[1]{\textcolor{purple}{\textbf{GZ: #1}}}

\DeclareMathOperator{\RY}{RY}
\DeclareMathOperator{\CRY}{CRY}
\DeclareMathOperator{\CNOT}{CNOT}

\DeclareMathOperator{\Diag}{Diag}

\DeclareMathOperator{\End}{End}

\newcommand*\bu{\boldsymbol{u}}
\newcommand*\ba{\boldsymbol{a}}
\newcommand*\bn{\boldsymbol{n}}
\newcommand*\bL{\boldsymbol{L}}
\newcommand*\bD{\boldsymbol{D}}
\newcommand*\bR{\boldsymbol{R}}

\newcommand*\bG{\boldsymbol{G}}

\newcommand*\bP{\boldsymbol{P}}

\newcommand*\bQ{\boldsymbol{Q}}

\newcommand*\bnu{\boldsymbol{\nu}}

\newcommand{\bbC}{\mathbb{C}}

\theoremstyle{plain}
\newtheorem{theorem}{Theorem}
\newtheorem{lemma}[theorem]{Lemma}
\newtheorem{definition}[theorem]{Definition}

\begin{document}

\title{Quantum Circuit for Non-Unitary Linear Transformation of Basis
Sets}

\author{Guorui Zhu}
\affiliation{School of Mathematical Sciences, Fudan University}
\author{Joel Bierman}
\affiliation{Electrical and Computer Engineering, North Carolina State
University}
\author{Jianfeng Lu}
\affiliation{Department of Mathematics, Department of Physics and
Department of Chemistry, Duke University}
\email[Corresponding author: ]{jianfeng@math.duke.edu}
\author{Yingzhou Li}
\affiliation{School of Mathematical Sciences, Shanghai Key Laboratory for
Contemporary Applied Mathematics, Fudan University} \affiliation{Key
Laboratory of Computational Physical Sciences, Ministry of Education}
\email[Corresponding author: ]{yingzhouli@fudan.edu.cn}

\begin{abstract}
    This paper introduces a novel approach to implementing non-unitary
    linear transformations of basis on quantum computational platforms, a
    significant leap beyond the conventional unitary methods. By
    integrating Singular Value Decomposition (SVD) into the process, the
    method achieves an operational depth of $O(n)$ with about $n$ ancilla
    qubits, enhancing the computational capabilities for analyzing
    fermionic systems. The non-unitarity of the transformation allows us
    to transform a wave function from one basis to another, which can span
    different spaces. By this trick, we can calculate the overlap of two
    wavefunctions that live in different (but non-distinct Hilbert
    subspaces) with different basis representations. This provides the
    opportunity to use state specific ansatzes to calculate different
    energy eigenstates under orbital-optimized settings and may improve
    the accuracy when computing the energies of multiple eigenstates
    simultaneously in VQE or other framework. It allows for a deeper
    exploration of complex quantum states and phenomena, expanding the
    practical applications of quantum computing in physics and chemistry. 
\end{abstract}

\maketitle

\section{Introduction}

Solving many-body Schr{\"o}dinger equation is one of the most promising
applications on quantum computers, especially in the era of noisy
intermediate-scale and early fault-tolerant quantum computing. Quantum
algorithms that could be applied to solve the many-body Schr{\"o}dinger
equation include but not limited to variational quantum
algorithms~(VQAs)~\cite{Peruzzo2014, Bauer2016, Kandala2017, Cerezo2021},
quantum phase estimation~(QPE)~\cite{Kitaev1995, Abrams1997, Abrams1998,
Nielsen2011}, quantum simulation~\cite{Georgescu2014}, etc. On digital
quantum computers, the many-body Schr{\"o}dinger equation first needs to
be discretized on a basis set, and then the discretized operator and
wavefunction need to be represented by quantum circuits and a quantum
state, respectively. Throughout the computation, when a change of basis
set is performed, i.e., a linear transformation of the basis set, we could
either recalculate coefficients on classical computer and restart the
quantum algorithm, or carry out the linear transformation via a quantum
circuit. Implementing quantum circuit for the linear transformation is
mandatory under various scenarios, e.g., calculating the inner product
between two many-body wavefunctions discretized by different basis sets.
When two basis sets are unitary linear transformations of each other, the
corresponding linear transformation is unitary and the quantum circuit has
been studied in~\cite{Wecker2015, kivlichan2018, Babbush2018}. When two
basis sets are not unitary linear transformations of each other, the
corresponding linear transformation is non-unitary, for which the quantum
circuit implementation is proposed by this paper. 

Unitary linear transformation of basis set plays an important role in
quantum physics and chemistry. Hartree-Fock method~\cite{Slater1930,
Slater1951, Roothaan1960, Vautherin1972, Baerends1973, Johnson2005,
Neese2009} looks for the optimal unitary linear transformation of basis
set such that the single Slater determinant minimizes the energy of the
many-body Hamiltonian operator. Unitary linear transformations of basis
set are also used widely in basis set optimization methods, e.g., natural
orbital rotation~\cite{Roothaan1960, Vautherin1972, Baerends1973,
Johnson2005, Neese2009}, complete active space self-consistent field
method~(CASSCF)~\cite{Siegbahn1980, Siegbahn1981, Roos1980, Knowles1985,
Zgid2008, Ghosh2008, Yanai2009, Olsen2011, Wouters2014, Freitag2017,
Smith2017, Sun2017, Freitag2019, Kreplin2019, Levine2020}, optimal orbital
full configuration interaction~(OptOrbFCI)~\cite{Li2020OptimalOS}, etc.
Such unitary linear transformations of basis sets are combined with
quantum algorithms for ground state calculation~\cite{Mizukami2020,
Tilly2021, Yalouz2021, Bierman2023} to boost the power of quantum computer
and pursuing the infinite basis set limit. The unitary linear basis
transformation can also be adopted in quantum algorithms for excited
states calculation~\cite{Tilly2021, Yalouz2021, Bierman2024} as long as
they use the same basis set to represent both ground state and low-lying
excited states of the system.

Several prior works studied the quantum circuit implementation of the
unitary linear transformation of basis set. The unitary coupled
cluster~(UCC) method~\cite{Yangchao2017, Taube2006, Joonho2018,
Romero2018, Anand2022}, particularly with single excitation operators, is
widely used as parameterized ansatz circuits for wavefunctions. UCC is
achieved by employing an operator $e^{\hat{T}}$, where $\hat{T}$ is
anti-Hermitian and represented as $\sum_{i < j} t_{ij}(\ba^\dag_i \ba_j -
\ba^\dag_j \ba_i)$. Trotter splitting is then adopted to approximate the
operator by a quantum circuit~\cite{Trotter1959, Suzuki1993, Hatano2005,
Vidal2003}. While this UCC quantum circuit is not designed for unitary
basis transformation, it could be used as a unitary transformation of the
basis set, which is mathematically guaranteed by Thouless
theorem~\cite{Thouless1960}. Some other methods are proposed when they
consider preparing the Slater determinant state by rotating the basis set.
\citet{Wecker2015} describe a procedure in which they use a quantum
circuit to take unitary transformation of a basis set and prepare
arbitrary Slater determinant state. Their circuit has no more than $n^2$
gates with arbitrary connectivity for $n$ being the size of the basis set.
\citet{Babbush2018} suggest employing the fermionic fast Fourier transform
to prepare a Slater determinant state in a plane wave basis. This approach
involves rotating the system from a plane wave dual basis, achieving a
depth of $O(n)$ while adhering to the planar lattice connectivity
constraints found in some existing superconducting quantum
platforms~\cite{Babbush2018}. The plane wave dual basis serves as a smooth
approximation to a grid structure, which has been explored to enhance the
efficiency of density functional
calculations~\cite{Skylaris2002,Skylaris2005}. In \cite{kivlichan2018}, an
efficient strategy is proposed to conduct the unitary linear
transformation on a quantum computer with linearly connected qubits, and
achieves a gate depth $\frac{n}{2}$, where the gate depth is counted as
the number of sequential Givens rotations. The strategy is a variant of
the QR decomposition based method of constructing single-particle unitary
transformations~\cite{Wecker2015,Maslov2007,Reck1994}.
\citet{kivlichan2018} organize the Givens rotations in QR decomposition in
a particular ordering to benefit from the parallelization on linearly
connected quantum computer and eliminates redundant rotations based on the
symmetry of the Hamiltonian. Some of these unitary linear transformation
methods~\cite{Babbush2018, kivlichan2018} target the quantum circuit
implementations of initial state preparations.

Non-unitary linear transformations of basis sets also play an important
role in quantum physics and chemistry for both ground state and excited
state computations. In ground state computations, for example,
\citet{Jimenez2012} use the non-unitary Thouless
theorem~\cite{Thouless1960} to rotate a orthogonal basis Hartree-Fock
state to a non-orthogonal basis Hartree-Fock state. \citet{Jimenez2012}
derive the formulas of the Hamiltonian matrix elements of this
non-orthogonal Hartree-Fock state. In the excited state computation, if
different basis sets~(either different basis sets or different
transformations of a basis set) are adopted for the ground state and
low-lying excited states, calculating the overlap between different states
involves a non-unitary linear transformation of basis set. Such a
state-specific rotation method for excited state computation could be
viewed as a multi-reference~(MR) method~\cite{Malmqvist1989, Pittner2009}.
Several studies have shown that the integration of state-specific methods
with CASSCF can yield higher accuracy, even with a reduced active
space~\cite{Kossoski2023, Marie2023, Yalouz2023, Saade2024}. Such an
approach is referred to as the state-specific CASSCF~(SS-CASSCF).
SS-CASSCF employs different reference states to generate configuration
interaction (CI) functions and incorporates an appropriate orthogonal
penalty into the objective function. The computation of this orthogonal
penalty requires computing the inner product between two states under
different bases. In general, if the states exhibit dense configurations,
the calculation of such inner products scales exponentially with respect
to the basis set size on classical computers. Computing such inner
products under special cases, \emph{e.g.} Hartree-Fock state or sparse
states, has been explored on classical computers~\cite{Malmqvist1989,
Granucci2001, Tapavicza2007, Mitric2008, Pittner2009, Likai2015,
Plasser2016}.

In this work, we propose a quantum circuit that performs the exact
non-unitary linear transformation of a basis set. The gate complexity of
the proposed quantum circuit scales polynomially in the basis set size,
more specifically scales quadratically in the basis set size, i.e.,
$O(n^2)$ for $n$ being the basis set size. The circuit depth scales
linearly in the basis set size. Combined with regular inner product
quantum circuit, we could evaluate the inner product of two states under
different basis sets with $O(n^2)$ gates and $O(n)$ depth on a quantum
computer. Our major contributions are summarized as follows.

\begin{enumerate}
    \item We rewrite the unitary linear transformation of basis set using
    wedge exterior notation, which can then be easily extended to the
    non-unitary case.

    \item A quantum circuit is proposed for non-unitary linear
    transformations of basis sets. The 1 and 0 singular values of the
    overlapping matrix between two basis sets are further compressed to
    reduce the gate complexity.

    \item Quantum circuits are proposed to evaluate the inner product of
    two states in different basis sets.
\end{enumerate}

The rest of the paper is organized as follows. In
\cref{sec:unitary_transformation}, we review the quantum circuit for
unitary linear transformations~\cite{kivlichan2018} and rewrite the
derivation of the unitary transformation operator in the language of exterior
algebra. \Cref{sec:nonunitary_transformation} proposes the novel quantum
circuit for performing non-unitary linear transformations of basis
set. The corresponding gate complexity and circuit depths are analyzed as
well. Based on the non-unitary linear transformation, quantum circuits
with depth $O(n)$ are proposed for evaluating the inner product of two
states under different basis sets in \cref{sec:inner_product}. Finally,
\cref{sec:summary} concludes the paper together with discussions on future
work.

\section{Unitary Linear Transformation}
\label{sec:unitary_transformation}

We review the quantum circuit for unitary linear transformation proposed
by \citet{kivlichan2018} and rewrite the derivation using exterior algebra
in this section.

The unitary linear transformation of basis set discussed
in~\cite{kivlichan2018} admits,
\begin{equation} \label{eq:basis_rotation}
    \ket{\phi_p} = \sum_{q=1}^n \ket{\psi_{q}} u_{qp}
\end{equation}
where $u$ is an $n\times n$ unitary matrix, ${\left\{ \ket{\psi_i}
\right\}}_{i=1}^n$ and ${\left\{ \ket{\phi_i} \right\}}_{i=1}^n $ are the
original and rotated orthonormal basis~(spin-orbitals) respectively, and
$n$ denotes the basis set size. The associated creation and annihilation
operators admit a similar transformation relationship,
\begin{equation*}
    \ba^\dag(\phi_p) = \sum_{q=1}^n \ba^\dag(\psi_q) u_{qp}
    \text{ and }
    \quad \ba(\phi_p) = \sum_{q=1}^n \ba(\psi_q) {(u_{qp})}^*,
\end{equation*}
where $\ba^{\dag}(\cdot)$ and $\ba(\cdot)$ are the creation and
annihilation operators associated with the given basis, and ${(u_{qp})}^*$
is the complex conjugate of $u_{qp}$.

Given the two one-body basis sets, $\{\ket{\psi_i}\}_{i=1}^n$ and
$\{\ket{\phi_i}\}_{i=1}^n$, we could generate two many-body basis sets to
represent the many-body states. For two one-body basis sets that are
related to each other by a unitary linear transformation, as in
\eqref{eq:basis_rotation}, their associated many-body basis sets are also
related to each other by a unitary linear transformation. The unitary
linear transformation in many-body space is characterized by the Thouless
theorem~\cite{Thouless1960}, which is equivalent to applying the operator,
\begin{equation}\label{eq:thouless_theorem}
     U(u;\psi) 
    = \exp({\sum_{p,q=1}^n{(\log u)}_{pq}\ba^\dag(\psi_p)
    \ba(\psi_q)})
\end{equation}
to the many-body states, where ${(\log u)}_{pq}$ denotes the $(p,q)$
element of the matrix $\log u$. The notation $\psi$ in $U(u;\psi)$ denotes
that all creation and annihilation operators correspond to the basis set
$\{\ket{\psi_i}\}_{i=1}^n$.

Now we give a definition of this $U(u;\psi)$ in the language of exterior
algebra. Given an $n$ dimensional Hilbert space $V$ with
$\{\ket{\psi_i}\}_{i=1}^n$ being its orthonormal basis set, the $k$-th
exterior powers is denoted as $\wedge^k V$. A basis set of $\wedge^k V$
admits,
\begin{equation} \label{eq:wedgekVbasis}
    \left\{
        \ket{\psi_{i_1}} \wedge \cdots \wedge \ket{\psi_{i_k}}
        \mid
        1 \leq i_1  < \cdots < i_k \leq n
    \right\}.
\end{equation}
The bases in \eqref{eq:wedgekVbasis} are orthonormal. For the electronic
many-body Schr{\"o}dinger equation, the $k$-particle states live in
$\wedge^k V$. The exterior algebra of $V$ is defined as, $\wedge V =
\oplus_{k \geq 0} \wedge^k V$. An orthonormal basis set of $\wedge V$
admits,
\begin{equation} \label{eq:wedgeVbasis}
    \left\{
        \ket{\psi_{i_1}} \wedge \cdots \wedge \ket{\psi_{i_k}}
        \mid
        1 \leq i_1  < \cdots < i_k \leq n \text{ and } k \geq 0
    \right\}.
\end{equation}
The space $\wedge V$ is also known as the Fock space in physics and
chemistry. In the following \cref{def:wedged_mapping}, we extend a linear
map $\bu: V \to V$ to a linear map $\wedge \bu : \wedge V \to \wedge V$.

\begin{definition} \label{def:wedged_mapping}
    Let $\bu : V \to V $ be a linear map. Define $\wedge \bu : \wedge V
    \to \wedge V$ as,
    \begin{equation*}
        \wedge \bu (\ket{\psi_{i_1}} \wedge \ket{\psi_{i_2}} \wedge \cdots
        \wedge \ket{\psi_{i_k}}) 
        = \bu \ket{\psi_{i_1}} \wedge \bu \ket{\psi_{i_2}} \wedge \cdots
        \wedge \bu \ket{\psi_{i_k}},
    \end{equation*}
    for $1 \leq i_1  < \cdots < i_k \leq n$ and $k \geq 0$. The linear map
    $\wedge\bu$ is called the wedged map of $\bu$.
\end{definition}

When the linear map $\bu$ is of form,
\begin{equation} \label{eq:bu}
    \bu = \sum_{i,j=1}^n u_{ij}\ket{\psi_i}\bra{\psi_j},
\end{equation}
the linear operator $U(u; \psi)$ as in \eqref{eq:thouless_theorem} is
consistent with $\wedge \bu$, i.e.,
\begin{equation*}
    U(u; \psi) = \wedge \bu.
\end{equation*}
\Cref{app:details_of_unitary_case} gives the detailed derivation. The
derivation gives a proof of Thouless theorem using the language of
exterior algebra, which is different from the BCH formula used in
\citet{kivlichan2018}. We find that the extending linear map $\wedge \bu$
to non-unitary transformation is straightforward, whereas extending the
operator $U(u; \psi)$ to non-unitary case is more difficult. Hence, the
linear map $\wedge \bu$ in exterior algebra will be used throughout the
rest paper mainly for non-unitary transformations.

An important property of $U(u; \psi)$ is the operator composition
property, which builds the foundation for~\cite{kivlichan2018}. Precisely,
for any two unitary matrices $u^1$ and $u^2$, the composition of $U(u^1;
\psi)$ and $U(u^2; \psi)$ is the operator of product of two matrices,
i.e.,
\begin{equation} \label{eq:matrix_homomorphism}
    U(u^1;\psi)U(u^2;\psi) = U(u^1u^2;\psi).
\end{equation}

\Cref{lem:wedge_homomorphism} gives an analog composition property for
wedged maps.

\begin{lemma} \label{lem:wedge_homomorphism}
    Suppose $\bu^1$ and $\bu^2$ are two linear maps $V \to V$, then
    the composition property holds for wedged maps,
    \begin{equation*}
        (\wedge \bu^1) (\wedge \bu^2) = \wedge(\bu^1 \bu^2).
    \end{equation*}
\end{lemma}

We emphasize that two linear maps $\bu^1$ and $\bu^2$ in
\cref{lem:wedge_homomorphism} are not necessary unitary.
\Cref{lem:wedge_homomorphism} holds for non-unitary linear maps as well.

Specifically, we consider two unitary maps defined as,
\begin{equation*}
    \bu^1 = \sum_{i,j=1}^n u^1_{ij} \ket{\psi_i}\bra{\psi_j}, \quad
    \bu^2 = \sum_{i,j=1}^n u^2_{ij} \ket{\psi_i}\bra{\psi_j}.
\end{equation*}
By the equivalence between $U(u; \psi)$ and $\wedge \bu$, we have $U(u^1;
\psi) = \wedge \bu^1$ and $U(u^2; \psi) = \wedge \bu^2$. Therefore, the
composition of two operators admits,
\begin{equation*}
    U(u^1u^2; \psi) = U(u^1; \psi) U(u^2; \psi)
    = (\wedge\bu^1) (\wedge\bu^2) = \wedge(\bu^1\bu^2).
\end{equation*}
Such a composition property agrees with the composition of maps $\bu^1$
and $\bu^2$,
\begin{equation*}
    \bu^1\bu^2 = \sum_{i,j=1}^n {(u^1u^2)}_{ij}\ket{\psi_i}\bra{\psi_j},
\end{equation*}
where $u^1 u^2$ denotes the product of two matrices.

Next, we discuss the quantum circuit implementation of $U(u; \psi)$. For
general unitary operator in exponent form,
e.g.,~\eqref{eq:thouless_theorem}, Trotterization is widely adopted in
quantum computing, especially in quantum simulation \cite{Seth1996,
Nielsen2011}. The Trotterization leads to many number of small time steps
and accumulation of truncation errors. For the particular unitary operator
$U(u; \psi)$ as in \eqref{eq:thouless_theorem}, \citet{kivlichan2018}
proposes a tailored quantum circuit based on Givens QR factorization of
$u$. In the original work~\cite{kivlichan2018}, real orthogonal matrix $u$
are considered in detail, and an expression for complex unitary $u$ is
provided without detail. In this paper, we revisit and rederive the
decomposition for complex unitary matrix $u$ directly. Notably, the final
expression is slightly different from that in~\cite{kivlichan2018}, which
should be due to typos therein.

The complex Givens rotation is composed of three parts: two phase
rotations and a regular real Givens rotation. The phase rotation acting on
the $q$-th row of a matrix with rotation angle $\phi$ is denoted as
$p_q(\phi) = \Diag\{1, \ldots, 1, e^{-\imath \phi}, 1 \ldots, 1\}$, i.e.,
the identity matrix except the $p$-th $1$ replaced by $e^{-\imath \phi}$.
The regular real Givens rotation by an angle $\theta$ between the $p$-th
and $q$-th rows of a matrix is denoted as $r_{pq}(\theta_{pq})$. Contrast
to the regular real Givens rotation, the complex Givens rotation first
multiplies the elements to be eliminated by complex signs, i.e., phase
rotations, to make them real, and then performs the real Givens rotation.
For example, let us consider the case using $u_{pi}$ to eliminate
$u_{qi}$, where both $u_{pi}$ and $u_{qi}$ are complex numbers. The phase
rotations, $p_p(\phi_p)$ and $p_q(\phi_q)$, are chosen such that $e^{-
\imath \phi_p} u_{pi} = |u_{pi}|$ and $e^{- \imath \phi_q} u_{qi} =
|u_{qi}|$, respectively. Then the real Givens rotation,
$r_{pq}(\theta_{pq})$, is constructed with,
\begin{equation*}
    \theta_{pq} = \arccos \frac{|u_{pi}|}{\sqrt{|u_{pi}|^2 + |u_{qi}|^2}}.
\end{equation*}
By these constructions, after we multiply $g_{pq} = r_{pq}(\theta_{pq})
p_q(\phi_q) p_p(\phi_p)$ to $u$, the $(q,i)$ entry is zeroed out. Using
the complex Givens rotation and regular Givens QR factorization procedure,
a unitary matrix $u$ is decomposed as,
\begin{equation} \label{eq:complex_qr}
    u = \prod_{pq}g_{pq}^* \cdot \prod_{i=1}^n p_i^*(\phi_i),
\end{equation}
where 
\begin{equation*}
    \begin{split}
        g^*_{pq} & = p_p(-\phi_p) p_q(-\phi_q) r_{pq}(-\theta_{pq}),
        \text{ and} \\
        p^*_i & = p_i(-\phi_i).
    \end{split}
\end{equation*}
Notice that there are $\frac{n(n-1)}{2}$ complex Givens rotations in the
first product of \eqref{eq:complex_qr}, and the ordering of $\{pq\}$ pairs
is not unique. \citet{kivlichan2018} propose a specific ordering of
$\{pq\}$ to maximize the parallelizability in applying these complex
Givens rotations on quantum computer.

By the definition of $U(u; \psi)$, we could define the operation of
$U(r_{pq}; \psi)$ and $U(p_p; \psi)$, and denote them, respectively, as
\begin{align}
    \bR_{pq}(\theta; \psi) &= U(r_{pq}(\theta);\psi), \text{ and}
    \label{eq:givens_rotation_gate}\\
    \bP_p(\phi; \psi) &= U(p_p(\phi); \psi). \label{eq:phase_gate}
\end{align}
The corresponding ``complex Givens rotation'' and its adjoint admit,
\begin{equation}
    \label{eq:complex_Givens_rotation_gate}
    \begin{split}
        \bG_{pq} & = U(g_{pq}) = \bR_{pq}(\theta_{pq}) \bP_q(\phi_q)
        \bP_p(\phi_p), \text{ and}\\
        \bG^*_{pq} & = U(g^*_{pq}) = \bP_p(-\phi_p) \bP_q(-\phi_q)
        \bR_{pq}(-\theta_{pq}),
    \end{split}
\end{equation}
where we omit $\psi$, and some $\theta_{pq}$, $\phi_p$, $\phi_q$ for
simplicity. Under these notations, the operation $U(u; \psi)$ could be
decomposed into many operations of $\bG_{pq}$ and $\bP_i$ for $u$ defined
in \eqref{eq:complex_qr},
\begin{equation} \label{eq:gate_list}
    U(u; \psi) = \prod_{pq} \bG^*_{pq} \prod_{i=1}^n\bP^*_i(\phi_i;\psi).
\end{equation}
Next we demonstrate that the decomposition \eqref{eq:gate_list} is
sufficiently simple and could be implemented using quantum gates directly.
By~\eqref{eq:thouless_theorem}, the explicit expressions of $\bP_{p} =
U(p_p(\phi);\psi)$ and $\bR_{pq} = U(r_{pq}(\theta);\psi)$ could be
written as
\begin{align}
    \bP_p(\phi;\psi) & = \exp(-\imath \phi_p \boldsymbol{n}(\psi_p)),
    \text{ and} \label{eq:explicit_expression_of_phase_gate}\\
    \bR_{pq}(\theta;\psi) & = \exp[{ \theta_{pq} (\ba^\dag(\psi_p)
    \ba(\psi_q) - \ba^\dag(\psi_q) \ba(\psi_p))}],
    \label{eq:explicit_expression_of_rotation_gate}
\end{align}
where the detailed derivations could be found
in~\cref{app:compile_details}.

Now, we turn to the quantum circuit designs for $U(u; \psi)$. To make our
discussion rigorous, we make a strict distinction between the quantum
state of the physical system being studied and the qubit state in the
encoded space of a quantum computer. They are linked to each other by
encoded mappings. Here the mapping is the Jordan-Wigner transformation as
in \cref{def:JW_mapping_maintext}.
\begin{definition} \label{def:JW_mapping_maintext}
    The Jordan-Wigner transformation of basis $\{\ket{f_i} \mid i = 1, 2,
    \ldots, n\}$ is denoted as $J_f$, such that:
    \begin{align*}
        J_f(\ket{f_{i_1}} \wedge \ket{f_{i_2}} \wedge \cdots \wedge
        \ket{f_{i_k}})
        = \ket{n_1 n_2 n_3 \cdots},
    \end{align*}
    where $n_1 n_2 n_3 \cdots$ is a bit string of length equals to the
    size of basis set $n$. And $n_i = 1$ if $i \in \{i_1, i_2, \ldots, i_k\}$,
    $n_i = 0$ otherwise.
\end{definition}
These basis vectors $\{\ket{f_i} \mid i = 1, 2, \ldots, n\}$ do
not need to be a physically meaningful basis, they can be arbitrary abstract
basis of any abstract vector space, such as $\ket{f_i}$ is the $i$-th unit
coordinate vector of $\bbC^n$. By this definition, the creation and
annihilation operators under Jordan-Wigner transformation are the same as
the usual definition, i.e.,

\begin{equation} \label{eq:JWadagaop}
    \begin{split}
        \ba_i^\dag & = J_f \ba^\dag(f_i) J_f^{-1} \\&=
        {Z}_1\otimes\cdots\otimes {Z}_{i-1} \otimes 
        (\ket{1} \bra{0}) \otimes {I}_{i+1} \otimes \cdots \otimes I_n,
        \text{ and}\\
        \ba_i & = J_f\ba(f_i)J_f^{-1} \\&=
        {Z}_1\otimes\cdots\otimes {Z}_{i-1} \otimes 
        (\ket{0} \bra{1}) \otimes {I}_{i+1} \otimes \cdots \otimes I_n.
    \end{split}
\end{equation}
Both ket-bra operations, $\ket{1}\bra{0}$ and $\ket{0}\bra{1}$,
are applied to the index $i$ qubit in the above definitions.
There is no basis notation in $\ba_i^\dag$ and $\ba_i$ because they are
operators on qubit states. Basis information are kept in the encoding
mapping $J_f$. This viewpoint is helpful in the section of inner-product
of two states under different bases in \cref{sec:inner_product}.

Consider a complex Givens rotation, it induces a rotation between rows $p$
and $q$ of the matrix $u$, allowing for the elimination of a single
element in one of those rows. If $p$ and $q$ are not adjacent, the image
under Jordan-Wigner transformation of the complex Givens rotation
$\bG_{pq}(\theta, \phi; \psi)$ will apply gates on all qubits between $p$
and $q$, which is due to the fact that the Jordan-Wigner transformation of
creators and annihilators are non-local. More explicitly, the complex
Givens rotation includes terms like $\ba^\dag(\psi_p) \ba(\psi_q)$, whose
Jordan-Wigner transformation admits,
\begin{align*}
    & J_\psi(\ba^\dag(\psi_p) \ba(\psi_q)) J_\psi^{-1} \\
    = & {Z}_1\otimes\cdots\otimes {Z}_{p-1} \otimes  \ket{1}\bra{0} \otimes \\
      &{Z}_{p+1}\otimes\cdots\otimes {Z}_{q-1} 
    \otimes \ket{0}\bra{1} \otimes {I}_{i+1} \otimes \cdots \otimes I_n.
\end{align*}
Therefore, if we use the regular Givens QR elimination order, i.e., using
the diagonal element to eliminate all the elements below it in its column,
it prevents us from applying complex Givens rotations in parallel. In the
worst case, it results in a quantum circuit of depth $O(n^2)$ for $n$
being the basis set size. An efficient Givens elimination order is
proposed in \citet{kivlichan2018} for qubits with linear connectivity. An
example with $n=5$ is depicted in \cref{fig:elimination_order}. Through
the elimination procedure, the complex Givens rotations are always applied
to neighboring rows and eliminate the elements on the second row. The
procedure boosts the parallelizability, and $O(n)$ elements, in the best
case, are simultaneously eliminated. Using the specific elimination order,
it results in a quantum circuit of depth $O(n)$.

\begin{figure}[htb]
    \centering
    $\begin{pmatrix}
        * & * & * & * & * \\
        4 & * & * & * & * \\
        3 & 5 & * & * & * \\
        2 & 4 & 6 & * & * \\
        1 & 3 & 5 & 7 & * \\
    \end{pmatrix}$
    \caption{Elimination order of elements via complex Givens rotations.
    The number describes the order that the element been eliminated.
    Elements with the same number mean that they could be eliminated
    simultaneously. Asterisks (*) mark all upper-diagonal
    elements.~\cite{kivlichan2018}}
    \label{fig:elimination_order}
\end{figure}

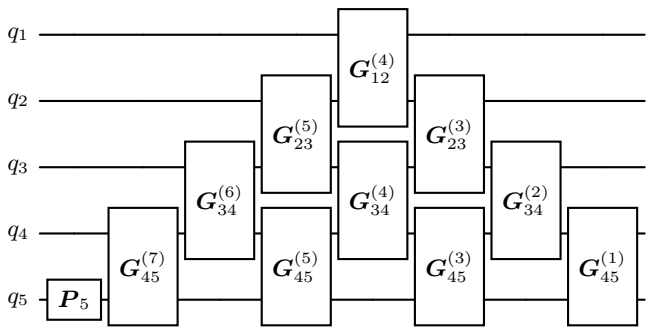
\begin{figure}[ht]
    \centering
    \begin{quantikz}[column sep= .1cm, row sep=.2cm]
        \lstick{$q_1$}& & & & & \gate[2]{\bG^{(4)}_{12}} & & & & \\
        \lstick{$q_2$}& & & & \gate[2]{\bG^{(5)}_{23}} & & \gate[2]{\bG^{(3)}_{23}} & & &\\
        \lstick{$q_3$}& & & \gate[2]{\bG^{(6)}_{34}} & & \gate[2]{\bG^{(4)}_{34}} & & \gate[2]{\bG^{(2)}_{34}} & &\\
        \lstick{$q_4$}& & \gate[2]{\bG^{(7)}_{45}} & & \gate[2]{\bG^{(5)}_{45}} & & \gate[2]{\bG^{(3)}_{45}} & & \gate[2]{\bG^{(1)}_{45}} &\\
        \lstick{$q_5$}&\gate{\bP_5}& & & & & & & & \\
    \end{quantikz}
    \caption{Quantum circuit for $J_\psi U(u;\psi)J_\psi^{-1}$. All gates
    are their complex conjugates.} \label{fig:quantum_circuit_of_unitary}
\end{figure}

We provide an observation that further simplifies the expression of $U(u;
\psi)$. After Givens QR upper triangularization, the resulting diagonal
matrix typically has all elements equal to $1$ except the last
bottom-right element. This is due to the phase rotations and unitarity of
$u$. Hence, the $U(u; \psi)$ expression~\eqref{eq:gate_list} could be
rephrased as 
\begin{equation} \label{eq:reduced_gate_list}
    U(u;\psi) = \prod_i\left[\prod_{q\in \text{Step}_i} (\bG^{(i)}_{q-1,q})^*\right] \cdot
    \bP^*_{n}(\phi_n; \psi).
\end{equation}
The upper index $i$ of the gate $\bG_{q-1,q}^{(i)}$ denotes the $i$-th step
of parallel complex Givens rotations and the set $\text{Step}_i$ is the
row indices of the elements eliminated in parallel at the $i$-th step.
Comparing to \eqref{eq:gate_list}, the product of phase operators is
simplified by a single phase operator acting on the $n$-th row, and
parallel sequence of the gates is also indicated. Now we give the quantum
circuit of $U(u; \psi)$. Under the Jordan-Wigner transformation with
respect to basis $\{\ket{\psi_i}\}_{i=1}^n$, the rotation
$\bR_{q-1,q}(\theta; \psi)$ is mainly represented by a $4 \times 4$ matrix
and $\bR_{q-1,q}(\theta; \psi)$ is encoded as,
\begin{equation} \label{eq:encoded_givens_rotation}
    \begin{split}
        & J_\psi \bR_{q-1,q}(\theta;\psi) J_\psi^{-1}\\
        = & I_1\otimes \cdots \otimes I_{q-2} \otimes 
        \begin{pmatrix}
            1&0&0&0\\
            0&\cos\theta&-\sin\theta&0\\
            0&\sin\theta&\cos\theta&0\\
            0&0&0&1
        \end{pmatrix} \otimes\\
        & I_{q+1}\otimes \cdots \otimes I_n,
    \end{split}
\end{equation}
where the middle $4 \times 4$ matrix can be expressed as the product of
three matrices as,
\begin{equation} \label{eq:raw_matrix_givens}
    \begin{pmatrix}
        1&0&0&0\\
        0&0&0&1\\
        0&0&1&0\\
        0&1&0&0
    \end{pmatrix}\cdot
    \begin{pmatrix}
        1&0&0&0\\
        0&1&0&0\\
        0&0&\cos\theta&\sin\theta\\
        0&0&-\sin\theta&\cos\theta
    \end{pmatrix}\cdot
    \begin{pmatrix}
        1&0&0&0\\
        0&0&0&1\\
        0&0&1&0\\
        0&1&0&0
    \end{pmatrix}.
\end{equation}
The first and third matrices in \eqref{eq:raw_matrix_givens} represent the
standard quantum $\CNOT$ gate, where the first qubit is the target and the
second qubit is the control. The second matrix in
\eqref{eq:raw_matrix_givens} represents the $\CRY(-2\theta)$ gate, where
the first qubit controls the $\RY(-2\theta)$ gate applied to the second
qubit. The matrix representation of $\RY(\theta)$ with angle $\theta$ is
given by
\begin{equation*}
    \RY(\theta) =
    \begin{pmatrix}
        \cos \frac{\theta}{2} & - \sin\frac{\theta}{2} \\
        \sin\frac{\theta}{2} & \cos\frac{\theta}{2}
    \end{pmatrix}.
\end{equation*}
As a result, if we denote the first line as the qubit $q-1$ and the second
as the qubit $q$, we have 
\begin{equation} \label{eq:encoded_givens_rotation_gate}
    J_\psi\bR_{q-1,q}(\theta;\psi)J_\psi^{-1} =
    \begin{quantikz}[column sep=.3cm]
        & \targ{} & \ctrl{1} & \targ{} & \qw \\
        & \ctrl{-1} & \gate{\RY(-2\theta)} & \ctrl{-1} & \qw
    \end{quantikz}.
\end{equation}
Then the phase operator is encoded as,
\begin{equation} \label{eq:encoded_phase}
    \begin{split}
        & J_\psi \bP_p(\phi;\psi)J_\psi^{-1} \\
        =\,& I_1\otimes \cdots \otimes I_{p-1}\otimes
        \begin{pmatrix}
            1&0\\
            0&e^{-i\phi_p}
        \end{pmatrix}
        \otimes I_{p+1} \otimes \cdots \otimes I_n.
    \end{split}
\end{equation}
In \cref{app:compile_details}, we give the detailed derivations for
\eqref{eq:encoded_givens_rotation} and \eqref{eq:encoded_phase}. The phase
operator is simply a phase gate acting on the $p$-th qubit. Combining the
rotation quantum circuit as in \eqref{eq:encoded_givens_rotation_gate}
together with that of the phase gate, we obtain the quantum circuit for
complex Givens rotations applied to adjacent rows, which is denoted as
$\bG_{q-1,q}^{(i)}$. The overall quantum circuit for $J_\psi
U(u;\psi)J_\psi^{-1}$ corresponding to a $5 \times 5$ unitary map $\bu$ is
given in \cref{fig:quantum_circuit_of_unitary}. The parallel steps are
consistent with that in \cref{fig:elimination_order}.

\emph{Remark.}
Comparing to the original work~\cite{kivlichan2018}, we extend the
discussion to complex unitary matrix in detail and provide the
corresponding quantum circuit. At the same time, we rewrite some
expressions using the exterior algebra notations, which serves as an
introduction of notations for our later sections.

\section{Non-Unitary Linear Transformation}
\label{sec:nonunitary_transformation}

We naturally raise a question: when $u$ is non-unitary, can we extend the
linear transformation in \cref{sec:unitary_transformation}? The answer is
yes, but the derivation is more complicated and the resulting quantum
circuit is about twice the depth of the unitary one. There are at least
two potential applications for the non-unitary linear transformation: a)
calculating the overlap between two states under different bases, which is
detailed in the next section; b) change of basis to a non-orthogonal one.

As we mentioned earlier, the Thouless theorem in
\eqref{eq:thouless_theorem} form is too difficult to be extended to
non-unitary case, which is mainly due to the logarithm of a non-unitary
matrix. Hence, we stick to exterior algebra representation of Thouless
theorem and all later derivations are in the language of exterior algebra.
We recall the notation we use, i.e., $U(u;\psi)$ and $\wedge \bu$. When
$u$ is a unitary matrix, two operators are equal $U(u; \psi) = \wedge
\bu$, for $U(u; \psi)$, $\wedge \bu$ and $\bu$ being defined in
\eqref{eq:thouless_theorem}, \cref{def:wedged_mapping}, and \eqref{eq:bu},
respectively. When $u$ is non-unitary, we use $\wedge \bu$ for the linear
map and avoid using $U(u; \psi)$.

In order to reduce the overall circuit depth for the non-unitary linear
transformation, our construction relies on the singular value
decomposition (SVD) of $u$ and hence its operator $\bu$. Let the SVD of a
non-unitary matrix $u$ be of form,
\begin{equation*}
    u = LDR,
\end{equation*}
where $L$ and $R$ are left and right singular vectors, $D =
\mathrm{diag}(\sigma_1, \dots, \sigma_n)$ is a diagonal matrix with
singular values in non-increasing ordering, i.e., $\sigma_1 \geq \cdots
\geq \sigma_n$. Without loss of generality, we further assume that $u$ is
a nonunitary matrix with 2-norm bounded by one, i.e., $1 \geq \sigma_1
\geq \dots \geq \sigma_n$. Singular vectors $L$ and $R$ are unitary
matrices. Let $\{\ket{\psi_i}\}_{i=1}^n$ be the set of underlying basis
set. The operator $\bu$ associated with $u$ also admits an operator SVD,
\begin{equation} \label{eq:SVD1}
    \bu = \bL \bD \bR,
\end{equation}
where $\bL$, $\bD$, and $\bR$ are associated with matrices $L$, $D$, and
$R$ respectively,
\begin{align*}
    \bL &= \sum_{i,j=1}^n L_{ij}\ket{\psi_i}\bra{\psi_j},\\
    \bD &= \sum_{i=1}^n \sigma_{i}\ket{\psi_i}\bra{\psi_i}, \text{ and}\\
    \bR &= \sum_{i,j=1}^n R_{ij}\ket{\psi_i}\bra{\psi_j}.\\
\end{align*}
By \cref{lem:wedge_homomorphism}, we have $\wedge \bu = (\wedge \bL)
(\wedge \bD) (\wedge \bR)$. For unitary matrices $L$ and $R$, the
operators $\wedge \bL = U(L;\psi)$ and $\wedge \bR = U(R;\psi)$ can be
implemented by quantum circuits as in \cref{sec:unitary_transformation},
e.g., analog quantum circuits of \cref{fig:quantum_circuit_of_unitary}. In
the following, we focus on the quantum circuit construction for the wedged
diagonal operator $\wedge \bD$.

We first rewrite $\wedge\bD$ as a composition of at most $n$ simple
operators, which are simple to be implemented by quantum gates. According
to \cref{def:wedged_mapping}, the action of $\wedge \bD$ on an exterior
algebra is defined as applying the operator $\bD$ to all one-body states,
as,
\begin{equation} \label{eq:action_wedge_d}
    \begin{split}
        & (\wedge\bD) \ket{\psi_{i_1}} \wedge \ket{\psi_{i_2}} \wedge \cdots
        \wedge \ket{\psi_{i_k}} \\
        =\,& \bD\ket{\psi_{i_1}} \wedge \bD\ket{\psi_{i_2}} \wedge \cdots
        \wedge \bD\ket{\psi_{i_k}}\\
        =\,& \sigma_{i_1} \sigma_{i_2} \cdots \sigma_{i_k} \ket{\psi_{i_1}}
        \wedge \ket{\psi_{i_2}}\wedge \cdots \wedge\ket{\psi_{i_k}}.
    \end{split}
\end{equation}
It is worth noting that this expression is closely related to the particle
number operator. It acts like a weighted number operator. More explicitly,
if the many-body state includes the state $\ket{\psi_i}$, the result of
the operation will give a factor $\sigma_i$. Thus, we define an operator
\begin{equation*}
    \bnu(\psi_j) = \boldsymbol{1} + (\sigma_j-1)\boldsymbol{n}(\psi_j),
\end{equation*}
which admits,
\begin{align*}
    & \bnu(\psi_j) (\ket{\psi_{i_1}} \wedge\ket{\psi_{i_2}}
    \wedge \cdots \wedge\ket{\psi_{i_k}}) \\
    = &
    \begin{cases}
        \sigma_j (\ket{\psi_{i_1}} \wedge \ket{\psi_{i_2}} \wedge \cdots
        \wedge \ket{\psi_{i_k}}), & j \in \{i_1, i_2, \ldots, i_k\}\\
        \ket{\psi_{i_1}} \wedge \ket{\psi_{i_2}} \wedge \cdots
        \wedge \ket{\psi_{i_k}}, & j \notin \{i_1, i_2, \ldots, i_k\}
    \end{cases}.
\end{align*}
Here $\bn(\psi_j) = \ba^\dagger(\psi_j) \ba(\psi_j)$ is the number
operator. Next, we justify that the wedge operator $\wedge \bD$ is
equivalent to the composition of $\bnu(\psi_j)$. For any basis
$\ket{\psi_{i_1}} \wedge \cdots \wedge \ket{\psi_{i_k}}$ in the exterior
algebra (Fock space), the action of the composition of $\bnu(\psi_j)$
admits,
\begin{align*}
    & \left(\prod_{j=1}^n \bnu(\psi_j)\right)
    \ket{\psi_{i_1}} \wedge \ket{\psi_{i_2}} \wedge \cdots
    \wedge \ket{\psi_{i_k}} \\
    =\,& \sigma_{i_1} \sigma_{i_2} \cdots \sigma_{i_k} \ket{\psi_{i_1}} \wedge
    \ket{\psi_{i_2}}\wedge \cdots \wedge\ket{\psi_{i_k}}
\end{align*}
which equals \eqref{eq:action_wedge_d}. Hence, we conclude that the
product operator $\prod_{j=1}^n \bnu(\psi_j)$ is another form of $\wedge
\bD$, i.e.,
\begin{equation*}
    \wedge\bD = \prod_{j=1}^n \bnu(\psi_j).
\end{equation*}

Now we move on to the quantum circuit implementation of composed operator
$\wedge\bD = \prod_{j=1}^n \bnu(\psi_j)$. Obviously, $\bnu(\psi_j)$ is not
a unitary operator, and hence, cannot be implemented by quantum circuits
directly. We introduce an ancilla qubit to encode the action of
$\bnu(\psi_j)$. In the context of the Jordan-Wigner transformation, the
particle number operator admits,
\begin{equation*}
    \boldsymbol{n}_j = I_1\otimes \cdots \otimes I_{j-1} \otimes
    \begin{pmatrix}
        0 & 0 \\
        0 & 1
    \end{pmatrix}
    \otimes I_{j+1} \otimes \cdots \otimes I_n.
\end{equation*}
We define an operator acting on qubit state as $\bnu_j$ admitting,
\begin{equation*}
    \begin{split}
        \bnu_j =\,& J_\psi \bnu(\psi_j) J_\psi^{-1} \\
        =\,& I_1\otimes \cdots \otimes I_{j-1} \otimes
        \begin{pmatrix}
            1 & 0 \\
            0 & \sigma_j
        \end{pmatrix}
        \otimes I_{j+1} \otimes \cdots \otimes I_n,
    \end{split}
\end{equation*}
where the operator norm is bounded by $1$ and so is $\sigma_j$.

We divide the discussion of quantum circuit implementation of $\bnu_j$
into three cases: $\sigma_j = 1$, $0 < \sigma_j < 1$, and $\sigma_j = 0$.
For simplicity, we drop the subscript $j$ in the following discussion.

When $\sigma = 1$, the corresponding operator $\bnu$ is an identity
operator, which does not require any quantum gates and any ancilla qubit.

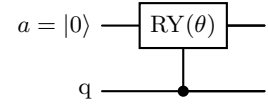
\begin{figure}[htb]
    \centering
    \begin{quantikz}
        \lstick{$a = \ket{0}$} & \gate{\RY(\theta)}  & \qw\\
        \lstick{q} & \ctrl{-1} & \qw\\
    \end{quantikz}
    \caption{Quantum circuit of embedded $2$-qubit gate. Ancilla qubit is
    denoted as $a$, which is initialized to $\ket{0}$ state. The second
    qubit $q$ is the working qubit which $\boldsymbol{\nu}$ is acted
    on.} \label{fig:circuit_of_entry_emedding}
\end{figure}

When $0 < \sigma < 1$, the corresponding operator $\bnu$ is nonunitary. To
incorporate this operation into a quantum circuit, we adopt an extra
ancilla qubit. The particular block encoding we use for the 2-by-2 matrix
in $\bnu$ admits,
\begin{align*}
    \tilde{\bnu} = &
    \begin{pmatrix}
        1&0&0&0\\
        0&\sigma&0&-\sqrt{1-\sigma^2}\\
        0&0&1&0\\
        0&\sqrt{1-\sigma^2}&0&\sigma
    \end{pmatrix} \\
    = &
    \begin{pmatrix}
        1&0&0&0\\
        0&\cos \frac{\theta}{2}&0&-\sin \frac{\theta}{2}\\
        0&0&1&0\\
        0&\sin \frac{\theta}{2}&0&\cos \frac{\theta}{2}
    \end{pmatrix},
\end{align*}
where the first qubit is the ancilla qubit and the second one is the
working qubit, and $\theta = 2 \arccos \sigma$. This encoded matrix
$\tilde{\bnu}$ on 2-qubit is the controlled-RY gate, denoted as
$\CRY(\theta)$. This controlled-RY gate applies a rotation around the
Y-axis by an angle $\theta$ to the ancilla qubit and controlled by the
working qubit. Denote $P_a$ as the projection operator that project the
ancilla qubit to $\ket{0}$ state, which could be implemented via a
measurement, then the construction $\tilde{\bnu}$ satisfies 
\begin{equation*}
    P_a \tilde{\bnu} \ket{0} \otimes \ket{\alpha} =
    \ket{0} \otimes \bnu \ket{\alpha}.
\end{equation*}
The quantum circuit for $\tilde{\bnu}$ is given in
\cref{fig:circuit_of_entry_emedding}. 

\begin{figure}[htb]
    \centering
    \begin{quantikz}[column sep=1cm, row sep=.5cm]
        \lstick{$a = \ket{0}$}& \gate{X} & \gate{X} &\\
        \lstick{$q_{r+1}$}& & \octrl{-1} &\\
        \lstick{$q_{r+2}$}& & \octrl{-1} &\\
        \vdots \\
        \lstick{$q_{n}$}  & & \octrl{-2} &\\
    \end{quantikz}
    \caption{Block encoding for all $\boldsymbol{\nu}_j$ with
    $\sigma_j=0$. Here the ancilla qubit $a$ is initialized to $\ket{0}$
    state. Other $q_{j}$ for $r < j \le n$ are working qubits
    corresponding to zero singular values of $u$.}
    \label{fig:circuit_of_0_singular_values}
\end{figure}
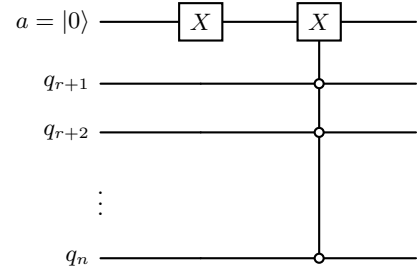

The last case is that $\sigma = 0$. The technique used for $0 < \sigma <
1$ case applies to the $\sigma = 0$ case as well. Then for each
$\tilde{\bnu}_j$ with singular value $\sigma_j=0$, we need an ancilla
qubit. Instead, we implement the quantum circuit for all $\sigma = 0$
together and use only one ancilla qubit to perform the block encoding.

Let $r = \arg\max_{i}\{\sigma_i>0\}$ be the rank of the nonunitary matrix
$u$. Then we have $\sigma_i > 0$ for $i = 1, \dots, r$ and $\sigma_i = 0$
for $i = r+1, \dots, n$. Noting that $\boldsymbol{\nu}_j$ with $\sigma_j=
0$ is the projection that projects qubit $q_j$ onto $\ket{0}$ state. Thus,
the composition of operators
\begin{equation*}
    \prod_{i = r + 1}^n \bnu_i
\end{equation*}
is the projection that projects qubits indexed from $r+1$ to $n$ to
$\ket{0 \cdots 0}$. It can be implemented as a $X$ gate combined with a
multi-open-controlled Toffoli gate. More specifically, given an ancilla
qubit at state $\ket{0}$ and working qubits $q_{r+1}, \dots, q_n$, we
first flip the ancilla qubit to state $\ket{1}$ via an $X$ gate. Then
targeting the ancilla qubit, we apply a multi-open-controlled Toffoli gate
controlled from all working qubits $q_{r+1}, \dots, q_n$. Open-controlled
gate means that the gate is applied only if the controlling qubits are in
state $\ket{0 \cdots 0}$. After these two gates, the quantum state will be
linear combination of
\begin{equation*}
    \ket{0}\ket{0\cdots 0} \quad \text{and} \quad
    \ket{1}\ket{0\cdots 0}_\perp,
\end{equation*}
where the first qubit is ancilla, the rest are working qubits $q_{r+1},
\dots, q_n$, and $\ket{0 \cdots 0}_\perp$ is a state perpendicular to
$\ket{0 \cdots 0}$. Hence, applying a measurement with selected result,
i.e., projecting the ancilla qubit to $\ket{0}$, will turn all working
qubits $q_{r+1}, \dots, q_n$ to state $\ket{0}$. Thus, such a quantum
circuit is the block encoding as we need. This scheme reduces the number
of ancilla qubits down to one for low-rank linear transformations $u$.
However, the actual quantum circuit depth in implementing the
multi-open-controlled Toffoli gate depends on the underlying quantum
computer hardware, which is beyond the scope of this paper.
\Cref{fig:circuit_of_0_singular_values} illustrates the detailed quantum
circuit for the case $\sigma = 0$. 

\begin{figure*}[htb]
    \centering
    \begin{quantikz}[column sep=.45cm, row sep=.7cm]
        \lstick{$\tilde{\sigma}_1=1,q_1$}   & \gate[8]{J_\psi U (R;\psi) J_\psi^{-1}} 
                                  &&&&        &        &        \gate[8]{J_\psi U (L;\psi) J_\psi^{-1}}
                                                                 &\\
        \lstick{$\tilde{\sigma}_2=1,q_2$}   &&&&&        &        &        &\\
        \lstick{$0<\tilde{\sigma}_3<1,q_3$} &&\ctrl{6}
                                   &&&        &        &        &\\
        \lstick{$0<\tilde{\sigma}_4<1,q_4$} &&&\ctrl{6}
                                    &&        &        &        &\\
        \lstick{$0<\tilde{\sigma}_5<1,q_5$} &&&&\ctrl{6}
                                     &        &        &        &\\
        \lstick{$\tilde{\sigma}_6=0, q_6$}  &&&&&&\octrl{6}&&\\
        \lstick{$\tilde{\sigma}_7=0, q_7$}  &&&&&&\octrl{5}&&\\
        \lstick{$\tilde{\sigma}_8=0, q_8$}  &&&&&&\octrl{4}&&\\
        \lstick{$a_3 = \ket{0}$} &&\gate{RY (\theta_3)}
                                   &&&        &        &        &\metercw[label style={inner sep=1pt}]{\ket{0}}\\
        \lstick{$a_4 = \ket{0}$} &&&\gate{RY (\theta_4)}
                                    &&        &        &        &\metercw[label style={inner sep=1pt}]{\ket{0}}\\
        \lstick{$a_5 = \ket{0}$} &&&&\gate{RY (\theta_5)}
                                     &        &        &        &\metercw[label style={inner sep=1pt}]{\ket{0}}\\
        \lstick{$a_6 = \ket{0}$} &&&&&\gate{X}&\gate{X}&        &\metercw[label style={inner sep=1pt}]{\ket{0}}\\
    \end{quantikz}
    \caption{An example for a nonunitary matrix $u$ of dimension $8$ with
    approximated singular values $\tilde{\sigma}_1 = \tilde{\sigma}_2 = 1$, $1 >
    \tilde{\sigma}_3 \ge \tilde{\sigma}_4 \ge \tilde{\sigma}_5 > 0$, and
    $\tilde{\sigma}_6 = \tilde{\sigma}_7 = \tilde{\sigma}_8 = 0$. This quantum
    circuit is the approximated block encoding of $\wedge\bu$.}
    \label{fig:quantum_circuit_of_nonunitary}
\end{figure*}
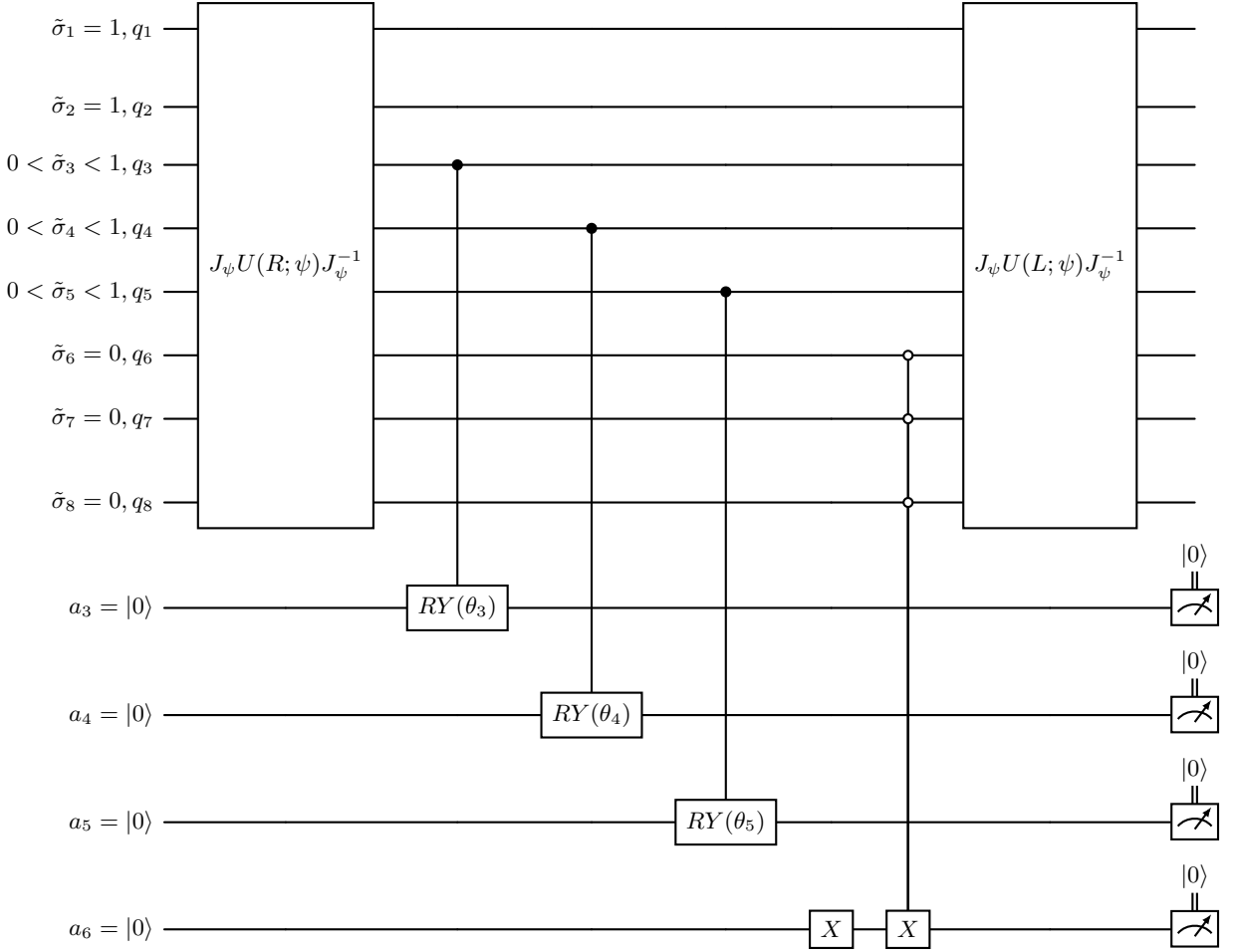

As discussed above, both cases $\sigma = 1$ and $\sigma = 0$ lead to more
quantum resource efficient circuit implementation than that of case $0 <
\sigma < 1$. Hence, naturally, we consider rounding up some large singular
values to one and truncating some small singular values to zero. Although
both rounding up and truncating reduce the quantum resource cost, both of
them would introduce approximation errors. Denote the modification on the
$j$-th singular values as $\epsilon_j$. The approximation error introduced
by both the rounding-up and truncating could be bounded as,
\begin{align*}
    \norm{\otimes_{j = 1}^n
    \begin{pmatrix}
        1 & 0\\
        0 & \sigma_j + \epsilon_j
    \end{pmatrix}
    - \otimes_{j = 1}^n
    \begin{pmatrix}
        1 & 0\\
        0 & \sigma_j
    \end{pmatrix}}
    \le \sum_{j = 1}^n \epsilon_j,
\end{align*}
where the matrix 2-norm is used. Let $\tilde{u}$ denote the matrix that
shares the same singular vectors as $u$, but has rounded up or truncated
singular values, i.e., $\tilde{u}_{ij} = \sum_{k=1}^n \tilde{\sigma}_k
L_{ik}R_{kj} = \sum_{k=1}^n (\sigma_k + \epsilon_k) L_{ik} R_{kj}$. In
this context, the error of approximated wedge operator can be bounded more
concisely as,
\begin{equation}\label{eq:error_bound_truncation}
    \norm{\wedge \bu - \wedge \tilde{\bu}} \le \sum_{j = 1}^n \epsilon_j.
\end{equation} Therefore, the overall approximation error could be
well-controlled if all $\epsilon_j$s are small.

Finally, we assemble all above techniques together and propose the quantum
circuit construction of a non-unitary linear transformation $\wedge \bu$.
To incorporate the rounding-up and truncating techniques, we introduce a
threshold $\varepsilon$. The quantum circuit for the non-unitary linear
transformation $\wedge \bu$ can be approximately implemented as follows.
\begin{enumerate}
    \item Calculate the SVD of $u$, i.e., $u_{ij} = \sum_{k=1}^n\sigma_k
    L_{ik}R_{kj}$, with singular values $\sigma_i$s in non-increasing
    ordering.

    \item Round up large singular values of $u$ and truncate small
    singular values of $u$, i.e.,
    \begin{equation*}
        \tilde{\sigma}_i =
        \begin{cases}
            1 & \sigma_i \geq 1 - \varepsilon \\
            \sigma_i & \varepsilon < \sigma_i < 1 - \varepsilon \\
            0 & \sigma_i \leq \varepsilon \\
        \end{cases}.
    \end{equation*}
    Let the number of $\tilde{\sigma}_i = 1$ be $s$ and the rank after
    truncation be $r$.

    \item Prepare $r-s+1$ ancilla qubits to state $\ket{0}$, denote them
    as $a_{s+1}, a_{s+1}, \dots, a_{r+1}$.

    \item Apply the unitary operator $J_\psi U(R;\psi)J_\psi^{-1}$ to the
    $n$ working qubits following the circuit construction in
    \cref{sec:unitary_transformation}.

    \item Apply controlled-RY gates controlling from $q_i$ targeting $a_i$
    for $i = s+1, \dots, r$.

    \item Apply an $X$ gate to flip ancilla qubit $a_{r+1}$ and then apply
    a multi-open-controlled Toffoli gate controlling from $q_{r+1}, \dots,
    q_{n}$ targeting $a_{r+1}$.

    \item Apply the unitary operator $J_\psi U(L;\psi)J_\psi^{-1}$ to the
    $n$ working qubits following the circuit construction in
    \cref{sec:unitary_transformation}.

    \item Project all ancilla qubits to $\ket{0}$ state via measurement
    and post-selection. 
\end{enumerate}
In \cref{fig:quantum_circuit_of_nonunitary}, we provide the quantum
circuit for a non-unitary matrix $u$ of dimension $8$. After the rounding-up
and truncating, the approximated non-unitary matrix has $2$ singular values
equal to $1$ and of rank $5$.

\noindent\emph{Remark.}
If we remove step $8$ and stop at step $7$, 
we actually obtain the quantum circuit corresponding
to the block encoding of $\wedge \bu$. Block-encoding is a standard framework used to embed non-unitary matrices into unitary matrices, allowing them to be implemented as quantum circuits along with some measurements~\cite{Gilyen2019}.
In the next section, we will discuss how to use the block-encoding we provided here in the context of calculating inner products.

The quantum circuit depth and the number of gates can be estimated based
on the matrix dimension $n$, approximated number of ones $s$ and
approximated rank $r$. The dimension of $L$ and $R$ are both $n$. And the
numbers of gates, as in \cref{sec:unitary_transformation} for the unitary
case, are bounded by $\frac{n^2}{2} + O(n)$ rotation gates
\eqref{eq:encoded_givens_rotation_gate} and $n^2 + O(n)$ phase gates. When
qubits are assumed to be linearly connected, the circuit depths for $L$
and $R$ are bounded by $O(n)$ times the rotation gate depth. The number of
gates for the $\prod_{k = 1}^{n}\bnu_k$ part depends on the gate counting
of the multi-open-controlled Toffoli gate. The quantum circuit for the
singular value part costs $r-s+2$ simple gates and one
multi-open-controlled Toffoli gate. The circuit depth is bounded by that
of the multi-open-controlled Toffoli gate. Applying the standard
decomposition as in~\cite{Nielsen2011}, an additional ancilla qubit is
used, and the quantum circuit depth is $O((n-r)^2)$ for $n-r$ being the
number of controlling qubits. Overall, the total quantum circuit depth is
bounded by $O({(n-r)}^2 + n)$.

\section{Inner product: An application}
\label{sec:inner_product}

We consider two sets of one-body bases, $\{ \ket{\psi_i} \}_{i=1}^n$ and
$\{ \ket{\phi_j} \}_{j=1}^n$. The inner products between all pairs of
bases $\ket{\psi_i}$ and $\ket{\phi_j}$ are available and denoted as
$u_{ij}$, i.e., $u_{ij} = \braket{\psi_i}{\phi_j}$, for $1 \le i, j \le
n$. For the sake of notation, we assume two bases are of the same
dimension $n$. Our proposed algorithm could be easily extended to two
bases of different dimensions. Two many-body states are denoted as
$\ket{\Psi}$ and $\ket{\Phi}$, i.e.,
\begin{align*}
    \ket{\Psi} & = \sum_{1 \le i_1 < i_2 < \cdots < i_k \le n}
    c_{i_1 i_2 \ldots i_k} \ket{\psi_{i_1}} \wedge \ket{\psi_{i_2}}
    \wedge \cdots \wedge \ket{\psi_{i_k}},\\
    \ket{\Phi} & = \sum_{1 \le i_1 < i_2 < \cdots < i_k \le n}
    c'_{i_1 i_2 \ldots i_k} \ket{\phi_{i_1}} \wedge \ket{\phi_{i_2}}
    \wedge \cdots \wedge \ket{\phi_{i_k}}.
\end{align*}
We consider the scenario such that both many-body states are already
encoded and prepared in quantum computer, where these states might be
generated from ansatz circuits used in VQE framework, from quantum
simulations, or initial state preparation methods. The encoded many-body
states on quantum computer are denoted as,
\begin{equation*}
    \ket{\Psi^q} = \sum_{I = 0}^{2^n-1} d_I \ket{I},
    \quad \text{and }
    \ket{\Phi^q} = \sum_{I = 0}^{2^n-1} d'_I\ket{I},
\end{equation*}
where $\ket{I}$ is a bit string representing states on quantum computer.
States $\ket{\Psi}$ and $\ket{\Psi^q}$ are connected via a quantum
encoding or transformation $\bQ_\psi$, where the subscript $\psi$
indicates that the encoding is related to basis
$\{\ket{\psi_i}\}_{i=1}^n$. Similarly, states $\ket{\Phi}$ and
$\ket{\Phi^q}$ are connected via $\bQ_\phi$. More precisely, the
connections admit,
\begin{equation*}
    \bQ_\psi \ket{\Psi} = \ket{\Psi^q},
    \quad \text{and }
    \bQ_\phi \ket{\Phi} = \ket{\Phi^q}.
\end{equation*}
When Jordan-Wigner encoding is adopted, $\bQ_\psi$ is the same as
$J_\psi$. Here, we use $\bQ_\psi$ as the encoding map to include other
encodings, e.g., parity encoding, Bravyi-Kitaev encoding, etc.

\begin{figure}[ht]
    \centering
    \begin{quantikz}
        \lstick{$\ket{0}$}      &\gate{H}    &\ctrl{1} &\gate{H}
                                                        &\metercw[label style={inner sep=1pt}]{\ket{0}}
                                                         &\\
        \lstick{$\ket{\Psi^q}$} &\qwbundle{n}&\swap{1} &&&\\
        \lstick{$\ket{\Phi^q}$} &\qwbundle{n}&\swap{-1}&&&\\
    \end{quantikz}
    \caption{Swap test circuit to calculate the modulus of the
    inner-product $\abs{\braket{\Phi^q}{\Psi^q}}$. The modulus square
    $\abs{\braket{\Phi^q}{\Psi^q}}^2$ equals to the probability of the
    measurement result being $\ket{0}$.} \label{fig:swaptest}
\end{figure}
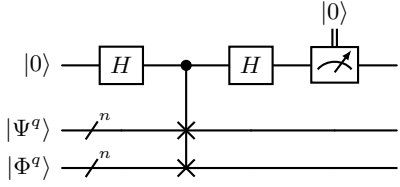

When two basis sets are the same and the encoding is also the same and
unitary, the inner product between two states $\ket{\Psi}$ and
$\ket{\Phi}$ is the same as that of two states on quantum computer, i.e.,
\begin{equation*}
    \braket{\Phi}{\Psi} = \braket{\Phi^q}{\Psi^q}.
\end{equation*}
Then, we could use many standard quantum circuits to evaluate it or its
modulus. A commonly used one is the swap test as in \cref{fig:swaptest}.
Other choices are also widely used in excited state
calculations~\cite{Bierman2022, Bierman2024}. However, when two states
$\ket{\Psi}$ and $\ket{\Phi}$ are under different basis sets, none of the
aforementioned quantum circuits for $\braket{\Phi^q}{\Psi^q}$ can be used
directly, due to the fact that $\braket{\Phi}{\Psi} \neq
\braket{\Phi^q}{\Psi^q}$. 

Notice that the inner product of two states is of a bilinear form. Without
loss of generality, we analyze the case that both many-body states are
many-body standard basis states, i.e.,
\begin{align*}
    \ket{\Psi} & = \ket{\psi_{i_1}} \wedge \ket{\psi_{i_2}} \wedge
    \cdots \wedge \ket{\psi_{i_k}},\\
    \ket{\Phi} & = \ket{\phi_{j_1}} \wedge \ket{\phi_{j_2}} \wedge
    \cdots \wedge \ket{\phi_{j_k}}.
\end{align*}
Further, we introduce an auxiliary basis $\{ \ket{\omega_i} \}_{i=1}^n$ to
facilitate the calculations. Note that the auxiliary basis will not affect
the results and its choice is arbitrary. We could choose it to be any
orthonormal basis in $\bbC^n$. The auxiliary space spanned by $\{
\ket{\omega_i} \}_{i=1}^n$ is denoted as $\Omega$. A linear map $\bu:
\Omega \to \Omega$ is defined as
\begin{equation} \label{eq:inner-u}
    \bu = \sum_{i,j = 1}^n \braket{\psi_i}{\phi_j}
    \ket{\omega_i}\bra{\omega_j}.
\end{equation}
Using the linear map $\bu$, the inner product of $\ket{\Psi}$ and
$\ket{\Phi}$ admits,
\begin{equation} \label{eq:PhiPsiinner1}
    \begin{split}
        \braket{\Phi}{\Psi}
        =\,& (\ket{\psi_{i_1}} \wedge \cdots \wedge \ket{\psi_{i_k}},
        \ket{\phi_{j_1}}\wedge \cdots \wedge\ket{\phi_{j_k}}) \\
        =\,& \det
        \begin{pmatrix}
            \braket{\psi_{i_1}}{\phi_{j_1}} & \cdots &
            \braket{\psi_{i_1}}{\phi_{j_k}} \\
            \vdots&\ddots&\vdots\\
            \braket{\psi_{i_k}}{\phi_{j_1}} & \cdots &
            \braket{\psi_{i_k}}{\phi_{j_k
            }} 
        \end{pmatrix}\\
        =\,& \det
        \begin{pmatrix}
            \bra{\omega_{i_1}} \bu \ket{\omega_{j_1}} & \cdots &
            \bra{\omega_{i_1}} \bu \ket{\omega_{j_k}} \\
            \vdots & \ddots & \vdots\\
            \bra{\omega_{i_k}} \bu \ket{\omega_{j_1}} & \cdots &
            \bra{\omega_{i_k}} \bu \ket{\omega_{j_k}} 
        \end{pmatrix}\\
        =\,& (\ket{\omega_{i_1}} \wedge \cdots \wedge \ket{\omega_{i_k}},
        \bu \ket{\omega_{j_1}} \wedge \cdots \wedge
        \bu \ket{\omega_{j_k}})\\
        =\,& (\ket{\omega_{i_1}} \wedge \cdots \wedge \ket{\omega_{i_k}},
        \wedge \bu
        \ket{\omega_{j_1}} \wedge \cdots \wedge \ket{\omega_{j_k}}).
    \end{split}
\end{equation}
Recall that $Q_\psi$, $Q_\phi$, and $Q_\omega$ are the same quantum
encoding technique applied to different basis sets. Hence, for two
many-body bases constructed by the same selection of one-body bases, their
quantum encoding map to the same state representation on quantum computer,
i.e., the same bit string representation. More precisely, for two
many-body bases, $\ket{\psi_{i_1}} \wedge \ket{\psi_{i_2}} \wedge \cdots
\wedge \ket{\psi_{i_k}}$ and $\ket{\omega_{i_1}} \wedge \ket{\omega_{i_2}}
\wedge \cdots \wedge \ket{\omega_{i_k}}$, which are constructed by
one-body bases of the same indices $i_1, i_2, \dots, i_k$, their quantum
encoded states are the same, i.e.,
\begin{equation*}
    \bQ_\psi (\ket{\psi_{i_1}} \wedge \ket{\psi_{i_2}} \wedge \cdots
    \wedge \ket{\psi_{i_k}}) = \bQ_\omega (\ket{\omega_{i_1}} \wedge
    \ket{\omega_{i_2}} \wedge \cdots \wedge \ket{\omega_{i_k}}).
\end{equation*}
Using such a relationship of quantum encoding, we continue deriving
\eqref{eq:PhiPsiinner1} as,
\begin{equation} \label{eq:PhiPsiinner2}
    \begin{split}
        \braket{\Phi}{\Psi}
        =\,&(\ket{\omega_{i_1}} \wedge \cdots \wedge \ket{\omega_{i_k}},
        \wedge \bu \ket{\omega_{j_1}} \wedge \cdots \wedge
        \ket{\omega_{j_k}}) \\
        =\,& (\bQ_\omega^{-1} \bQ_\omega \ket{\omega_{i_1}} \wedge \cdots
        \wedge \ket{\omega_{i_k}}, \\
        & (\wedge \bu) \bQ_\omega^{-1} \bQ_\omega
        \ket{\omega_{j_1}} \wedge \cdots \wedge \ket{\omega_{j_k}}) \\
        =\,& (\bQ_\omega^{-1} \bQ_\psi \ket{\psi_{i_1}} \wedge \cdots \wedge
        \ket{\psi_{i_k}}, \\
        & (\wedge \bu) \bQ_\omega^{-1} \bQ_\phi
        \ket{\phi_{j_1}} \wedge \cdots \wedge \ket{\phi_{j_k}}) \\
        =\,& (\bQ_\omega^{-1} \ket{\Psi^q},
        (\wedge \bu) \bQ_\omega^{-1} \ket{\Phi^q}).
    \end{split}
\end{equation}
Notice that the quantum encoding map $\bQ_\omega$ preserves the
inner-product, i.e.,
\begin{align*}
    & (\bQ_\omega \ket{\omega_{i_1}} \wedge \cdots
    \wedge \ket{\omega_{i_k}}, \bQ_\omega \ket{\omega_{j_1}} \wedge \cdots
    \wedge \ket{\omega_{j_k}}) \\
    =\,& (\ket{\omega_{i_1}} \wedge \cdots \wedge \ket{\omega_{i_k}},
    \ket{\omega_{j_1}} \wedge \cdots \wedge \ket{\omega_{j_k}}),
\end{align*}
for any $\omega$, and $i_1, \dots, i_k$, and $j_1, \dots, j_k$. Therefore,
in the last expression in \eqref{eq:PhiPsiinner2}, we could multiply both
side by $\bQ_\omega$ and obtain the definition of operator $\Xi(\psi, \phi)$,
\begin{equation} \label{eq:PhiPsiinner3}
    \braket{\Phi}{\Psi}
    = (\ket{\Psi^q}, \bQ_\omega (\wedge \bu) \bQ_\omega^{-1}
    \ket{\Phi^q})
    = \bra{\Psi^q} \Xi(\psi, \phi) \ket{\Phi^q},
\end{equation}
where
\begin{equation*}
    \Xi(\psi,\phi) = \bQ_\omega (\wedge \bu) \bQ_\omega^{-1}.
\end{equation*}
In \eqref{eq:PhiPsiinner3}, both $\ket{\Phi^q}$ and $\ket{\Psi^q}$ are
quantum encoded states, which are constructed by their own quantum
circuits. The remaining task in evaluating the inner product
$\braket{\Phi}{\Psi}$ is to construct $\Xi(\psi, \phi)$ as a quantum
circuit. From now on, we take $\bQ_\omega$ here as the Jordan-Wigner
mapping $J_\omega$ such that the quantum circuits in
\cref{sec:unitary_transformation} and \cref{sec:nonunitary_transformation}
could be reused directly. Then, the quantum circuit for $\Xi(\psi, \phi)$
is exactly the same as that proposed in
\cref{sec:nonunitary_transformation} with $u_{ij} =
\braket{\psi_i}{\phi_j}$ and $\{\omega_i\}_{i=1}^n$ being the underlying
basis set. More precisely, incorporating the quantum circuit for
$\Xi(\psi, \phi)$ as illustrated in
\cref{fig:quantum_circuit_of_nonunitary} and projecting all ancilla qubits
to state $\ket{0}$, we could obtain the quantum state $\Xi(\psi,
\phi)\ket{\Phi^q}$ on quantum computer. For a better understanding and
comparison between different inner product quantum circuits, we denote
$\tilde{\Xi}(\psi,\phi)$ as the block encoding of $\Xi(\psi,\phi)$. For
example, for $\Xi(\psi, \phi)$ in
\cref{fig:quantum_circuit_of_nonunitary}, $\tilde{\Xi}(\psi,\phi)$ is the
quantum circuit therein without projecting ancilla qubits to $\ket{0}$.

\begin{figure}[ht]
    \centering
    \begin{quantikz}
        \lstick{$\ket{0}$}      &\gate{H}      &&\ctrl{1}&\gate{H}
                                                        &\metercw[label style={inner sep=1pt}]{\ket{0}}&
                                                            \\
        \lstick{$\ket{\Psi^q}$} &\qwbundle{n}  &\gate[2]{\tilde{\Xi} (\psi,\phi)}
                                                &\swap{2} &&&\\
        \lstick{$\ket{0_a}$}    &\qwbundle{n_a}&&&&\metercw[label style={inner sep=1pt}]{\ket{0_a}}&\\
        \lstick{$\ket{\Phi^q}$} &\qwbundle{n}  &&\swap{-2}&&&\\
    \end{quantikz}
    \caption{Swap test circuit to calculate the inner-product
    $\abs{\braket{\Phi}{\Psi}}$. The absolute value square
    $\abs{\braket{\Phi}{\Psi}}^2$ is the probability of all the
    measurement being $\ket{0}$.}
    \label{fig:simplified_swaptest_no_ancilla}
\end{figure}
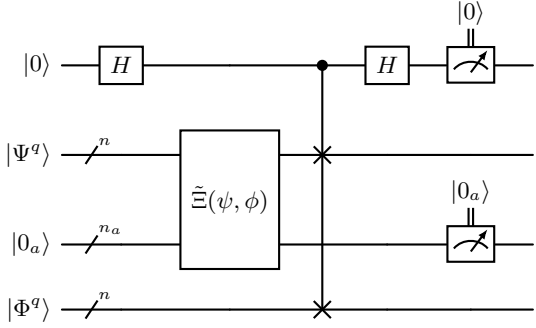

Combining the quantum circuit for $\Xi(\psi,\phi)$ and swap test as in
\cref{fig:swaptest}, we obtain our first quantum circuit for
$\abs{\braket{\Psi}{\Phi}}$ as in
\cref{fig:simplified_swaptest_no_ancilla}. Notice that in
\cref{fig:simplified_swaptest_no_ancilla}, the quantum circuit is
constructed using $\tilde{\Xi}(\psi,\phi)$ and all measurements are
postponed to the end and been measured simultaneously. This quantum
circuit only require one extra ancilla qubit for inner product. Instead,
we need to measure all ancilla qubits, including those from
$\Xi(\psi,\phi)$ and the one from swap test. The probability of $a+1$
ancilla qubits in $\ket{0}$ leads to the absolute value of the desired
inner product, $\abs{\braket{\Psi}{\Phi}}$. Slightly modified quantum
circuits could be used to evaluate the real and imaginary part of
$\braket{\Psi}{\Phi}$, and lead to the value of the desired inner product.

The quantum circuit in \cref{fig:simplified_swaptest_no_ancilla} evaluates
the absolute value of the inner product based on the following equation,
\begin{equation} \label{eq:inner_prod_qc}
    \braket{\Psi}{\Phi} = \bra{\Psi^q} \Xi(\psi,\phi) \ket{\Phi^q}
    = \bra{\Psi^q}\bra{0_a} \tilde{\Xi}(\psi,\phi) \ket{\Phi^q}\ket{0_a},
\end{equation}
where $\ket{0_a}$ denotes the ancilla qubits in the block encoding of
$\Xi(\psi, \phi)$. In \cref{fig:simplified_swaptest_no_ancilla}, the inner
product with $\bra{0_a}$ is carried out by the measurement projection and
the inner product with $\bra{\Psi^q}$ is carried out by the swap test.

\begin{figure}[ht]
    \centering
    \begin{quantikz}
        \lstick{$\ket{0}$}      &\gate{H}      &&\ctrl{1}&\ctrl{2}&\gate{H}
                                                        &\metercw[label style={inner sep=1pt}]{\ket{0}}
                                                            &\\
        \lstick{$\ket{\Psi^q}$} &\qwbundle{n}  &\gate[2]{\tilde{\Xi} (\psi,\phi)}
                                                &\swap{2} &&&&\\
        \lstick{$\ket{0_a}$}    &\qwbundle{n_a}&&&\octrl{2}&&&\\
        \lstick{$\ket{\Phi^q}$} &\qwbundle{n}  &&\swap{-2}&&&&\\
        \lstick{$\ket{0}$}      &              &&&\gate{X}&&&\\
    \end{quantikz}
    \caption{Alternative swap test circuit to calculate the inner-product
    $\abs{\braket{\Phi}{\Psi}}$. The absolute value square
    $\abs{\braket{\Phi}{\Psi}}^2$ is the probability of the ancilla
    measurement being $\ket{0}$.}
    \label{fig:simplified_swaptest}
\end{figure}
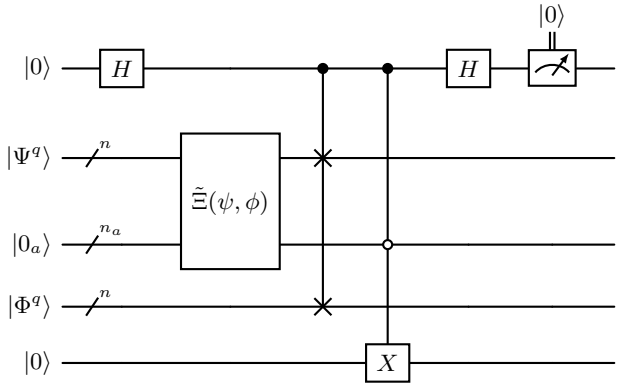

In some cases, measurements are not favored, e.g., the measurement errors
are large. Instead of the quantum circuit in
\cref{fig:simplified_swaptest_no_ancilla}, we could propose other quantum
circuits to evaluate the inner product as well.

An alternative quantum circuit for the inner product is to evaluate both
inner products with $\bra{0_a}$ and $\bra{\Psi^q}$ in
\eqref{eq:inner_prod_qc} using swap test. In this case, only one
measurement is needed, while extra $a$ ancilla qubits are required for
$\bra{0_a}$. In fact, we can simplify such a quantum circuit, replacing
the extra $a$ ancilla qubits by a single ancilla qubit and replacing the
controlled swap gates between ancilla qubits by a multi-controlled NOT
gate. The simplified quantum circuit is given in
\cref{fig:simplified_swaptest}.

\begin{figure}[ht]
    \centering
    \begin{quantikz}
        \lstick{$\ket{0}$}   &\gate{H}    &\ctrl{1}&\ctrl{1}&\ctrl{1}&\gate{H}
                                                        &\metercw[label style={inner sep=1pt}]{\ket{0}}
                                                            \\
        \lstick{$\ket{0_n}$} &\qwbundle{n}&\gate{U_\Phi}&\gate[2]{\tilde{\Xi}(\psi,\phi)}&\gate{U_\Psi^\dag}&&  \\
        \lstick{$\ket{0_a}$} &\qwbundle{a}&&&&&  \\
    \end{quantikz}
    \caption{Hadamard test circuit to calculate $\Re\braket{\Psi}{\Phi}$,
    which could be inferred from the probabilities of the measurement
    result.}\label{fig:hadamard_test}
\end{figure}
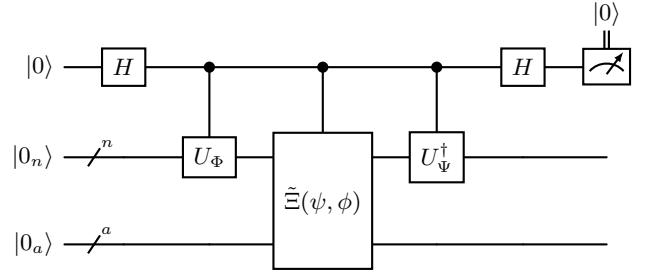

Another alternative quantum circuit for the inner product is based on the
Hadamard test. Here we consider a slightly different setting. Suppose that
we have quantum circuits $U_\Psi$ and $U_\Phi$ for preparing states
$\ket{\Psi^q}$ and $\ket{\Phi^q}$, respectively, i.e.,
\begin{equation*}
    \ket{\Psi^q} = U_\Psi\ket{0_n}, \text{ and } \ket{\Phi^q}
    = U_\Phi \ket{0_n}.
\end{equation*}
Under the VQE framework, $U_\Psi$ and $U_\Phi$ are known from
parameterized ansatz quantum circuits. Then, the inner product admits,
\begin{align*}
    \braket{\Phi}{\Psi}
    & = \bra{\Psi^q} \bra{0_a} \tilde{\Xi}(\psi,\phi)
    \ket{\Phi^q} \ket{0_a} \\
    & = \bra{0_{n+a}} (U_\Psi^\dag \otimes I_a) \tilde{\Xi}(\psi,\phi)
    (U_\Phi\otimes I_a) \ket{0_{n+a}}.
\end{align*}
The quantum circuit to evaluate the real part of $\braket{\Phi}{\Psi}$ is
given in \cref{fig:hadamard_test}, where the controlled-$(U_\Psi^\dag
\otimes I_a)$ and controlled-$(U_\Phi\otimes I_a)$ are simplified by the
controlled-$U_\Psi^\dag$ and controlled-$U_\Phi$ respectively. A simple
modification leads to the quantum circuit for the imaginary part of
$\braket{\Psi}{\Phi}$. When the adjoint of the ansatz circuit is not
favored, inner product quantum circuits proposed in \cite{Bierman2022}
could be used as an alternative to the Hadamard test.

\section{Summary}
\label{sec:summary}

This paper proposes a novel quantum circuit design for non-unitary linear
transformations of basis sets. The non-unitary linear transformations of
basis sets could be used in many practical scenarios, including but not
limited to the change of basis in initial state preparation, evaluating
the overlapping between two states under different basis sets in excited
state calculation under the VQE framework, etc.

Let $\bu$ be a non-unitary linear transformation of one-body basis sets,
$\bu: V \to V$ for $V$ being the space spanned by one-body bases. The
designed quantum circuit implements the wedged map on the exterior algebra
of $V$ (Fock Space), i.e., $\wedge \bu : \wedge V \to \wedge V$. To reduce
the overall circuit complexity, we first calculate an SVD of $\bu$. The
left and right singular vector operators of $\bu$ are unitary and are
implemented by the unitary linear transformation proposed in
\cite{kivlichan2018}, which is reviewed in
\cref{sec:unitary_transformation}. The quantum circuit for the singular
values of $\bu$ are detailed in \cref{sec:nonunitary_transformation}.
Block encodings are applied to all singular values strictly between zero
and one. The same block encoding could be applied to zero singular value
as well. Instead, we propose a quantum circuit to encoding all zero
singular values together with only one extra ancilla qubit. For singular
values that are sufficiently close to one or zero, we further round big
ones up to one and truncate small ones to zero. After the rounding-up and
truncating, the quantum circuit complexity is reduced while the
approximation errors are well-controlled. As a result, the proposed
quantum circuit achieves a depth of $O(n)$, for $n$ being the size of the
basis set. The extra required number of ancilla qubits is the number of
singular values strictly between zero and one.

Using the proposed quantum circuit for the non-unitary linear
transformation, we further provide quantum circuit evaluating the inner
product of two many-body states under different basis sets in
\cref{sec:inner_product}. Three quantum circuits for the inner product
evaluation are proposed with various number of ancilla qubits and
measurements. Combining the inner product quantum circuits with the
excited state calculation under VQE framework, we could immediately apply
different basis optimization for different many-body states, which
implement the state-specific CASSCF/OptOrbFCI on quantum computer.

\bibliography{reference}

\appendix

\section{Details of Unitary Case}
\label{app:details_of_unitary_case}

This section gives another proof of Thouless theorem using exterior
algebra. The proof offers a more intuitive understanding of the
construction process for the non-unitary linear transformations.
Additionally, it shows that the wedged map $\wedge \bu$ defined
in~\cref{def:wedged_mapping} is consistent $U(u; \psi)$ in Thouless
theorem and the work~\citet{kivlichan2018}.

Let $V$ be a finite-dimensional Hilbert spaces of dimension $n$. Operator
$\bu : V \to V$ is unitary, and admits \eqref{eq:bu}. Notice that $\bu$ is
unitary and is a finite dimensional normal operator. The
eigendecomposition of $\bu$ admits,
\begin{equation*}
    \bu = \sum_{k=1}^n e^{\imath \phi_k}\ket{\sigma_k}\bra{\sigma_k},
\end{equation*}
for $\ket{\sigma_k}$ being its eigenstate associated with eigenvalue
$e^{\imath \phi_k}$. The eigenstates of $\bu$,
$\{\ket{\sigma_k}\}_{k=1}^n$ are orthonormal bases of $V$. Hence, the set
of many-body states, $\{\ket{\sigma_{i_1}} \wedge \cdots \wedge
\ket{\sigma_k} \mid 1 \le i_1 < \cdots < i_k \le n, \text{ and } 1 \le k
\le n\}$ is also a basis of $\wedge V$.

Next, we derive the application of $\wedge \bu$ on bases of $\wedge V$.
By the definition of $\wedge \bu$, we have
\begin{equation} \label{eq:wedge_bu_num}
    \wedge \bu \ket{\sigma_{i_1}} \wedge \cdots \wedge \ket{\sigma_{i_k}}
    = e^{\sum_{k=1}^n \imath \phi_k n_k} \ket{\sigma_{i_1}} \wedge
    \cdots \wedge \ket{\sigma_{i_k}},
\end{equation}
for all $1 \leq i_1 < \cdots < i_k \leq n$, and $1 \leq k \leq n$, where
$n_k$ is defined as,
\begin{equation*}
    n_k =
    \begin{cases}
        0 & k \not\in \{i_1, \dots, i_k\},\\
        1 & k \in \{i_1, \dots, i_k\}.
    \end{cases}
\end{equation*}
Notice that $n_k$ plays a similar role as the particle number operator
$\bn(\sigma_k)$. The right-hand-side of \eqref{eq:wedge_bu_num} equals to
the application of $\prod_{k=1}^n e^{ \imath \phi_k \bn(\nu_k)}$ to
$\ket{\sigma_{i_1}} \wedge \cdots \wedge \ket{\sigma_{i_k}}$. Further, we
know that all particle number operators commute with each other. Hence, we
obtain,
\begin{align*}
    & \wedge \bu \ket{\sigma_{i_1}} \wedge \cdots \wedge \ket{\sigma_{i_k}} \\
    =\,& e^{\sum_{k=1}^n \imath \phi_k n_k} \ket{\sigma_{i_1}} \wedge
    \cdots \wedge \ket{\sigma_{i_k}} \\
    =\,& e^{\sum_{k=1}^n i\phi_k \bn(\sigma_k)} \ket{\sigma_{i_1}} \wedge
    \cdots \wedge \ket{\sigma_{i_k}},
\end{align*}
for all $1 \leq i_1 < \cdots < i_k \leq n$, and $1 \leq k \leq n$. We
conclude that $\wedge \bu = e^{\sum_{k=1}^n \imath \phi_k \bn(\sigma_k)}$.

Comparing to the operator $U(u; \psi)$, the basis in the above derivation
is $\{\sigma_k\}_{k=1}^n$, which is different from $\{\psi_k\}_{k=1}^n$ in
$U(u; \psi)$. Since both $\{\sigma_k\}_{k=1}^n$ and $\{\psi_k\}_{k=1}^n$
are bases of $V$, the basis transformation between these two is unitary.
As derived in \cite{Li2020OptimalOS}, the transformation of creators and
annihilators for two basis sets are
\begin{align*}
    \ba^\dag(\sigma_k) &= \sum_{m=1}^n \braket{\psi_m}{\sigma_k}
    \ba^\dag(\psi_m), \text{ and}\\
    \ba(\sigma_k) &= \sum_{m=1}^n \braket{\sigma_k}{\psi_m} \ba(\psi_m).
\end{align*}
The exponent in $e^{\sum_{k=1}^n \imath \phi_k \bn(\sigma_k)} = \wedge
\bu$ could be rewritten as,
\begin{align*}
    & \sum_{\ell=1}^n \imath \phi_\ell \bn(\sigma_\ell) \\
    =\,& \sum_{\ell=1}^n \imath \phi_\ell \ba^\dag(\sigma_\ell)
    \ba(\sigma_\ell) \\
    =\,& \sum_{\ell=1}^n \imath \phi_\ell
    \sum_{p=1}^n \braket{\psi_p}{\sigma_\ell} \ba^\dag(\psi_p)
    \sum_{q=1}^n \braket{\sigma_\ell}{\psi_q} \ba(\psi_q) \\
    =\,& \sum_{p=1}^n \sum_{q=1}^n \bra{\psi_p}
    \left( \sum_{\ell=1}^n \imath \phi_\ell \ket{\sigma_\ell}
    \bra{\sigma_\ell} \right) \ket{\psi_q} \ba^\dag(\psi_p) \ba(\psi_q) \\
    =\,& \sum_{p,q=1}^n \bra{\psi_p}(\log \bu)\ket{\psi_q}
    \ba^\dag(\psi_p) \ba(\psi_q).
\end{align*}
Recall the definition of operator $\bu$ under basis $\{\psi_k\}_{k=1}^n$
as in \eqref{eq:bu}. Applying the logarithm matrix function to $\bu$ leads to
\begin{equation} \label{eq:logu}
    \bra{\psi_p} \log\bu \ket{\psi_q} = {(\log u)}_{pq}.
\end{equation}    
Substituting \eqref{eq:logu} into the above exponent expression, we prove
the Thouless theorem using exterior algebra,
\begin{equation*}
    \wedge \bu = \exp({\sum_{p,q=1}^n {(\log u)}_{pq} \ba^\dag(\psi_p)
    \ba(\psi_q)}) = U(u;\psi).
\end{equation*}

\section{Complex Givens Rotation Operator}
\label{app:compile_details}

This section first derives \eqref{eq:explicit_expression_of_phase_gate}
and \eqref{eq:explicit_expression_of_rotation_gate}, and then
\eqref{eq:encoded_phase} and \eqref{eq:encoded_givens_rotation} in detail.

We first give the derivation of
\eqref{eq:explicit_expression_of_phase_gate}. The phase matrix $p_p(\phi)$
multiplies the complex sign $e^{-\imath \phi_p}$ to the $p$-th row of a
matrix. The matrix $p_p(\phi)$ is of form,
\begin{equation*}
        \Diag \{1, \ldots, 1, e^{- \imath \phi_p}, 1, \ldots ,1\}.
\end{equation*}
The logarithm of $p_p(\phi)$ is
\begin{equation*}
    \log p_{p}(\phi) = - \imath \phi_p E_{pp},
\end{equation*}
where $E_{pp}$ is a zero matrix with a single one at the $(p,p)$-th
position. Substituting the logarithm of $p_p(\phi)$ into $U(p_p(\phi);
\psi)$, we obtain,
\begin{align*}
    & U(p_{p}(\phi); \psi) \\
    =\,& \exp({- \imath \phi_p \sum_{i,j=1}^n {(E_{pp})}_{ij}
    \ba^\dag(\psi_i) \ba(\psi_j)}) \\
    =\,& \exp(- \imath \phi_{p} \bn(\psi_p)),
\end{align*}
which proves \eqref{eq:explicit_expression_of_phase_gate}.

Then we give the derivation of
\eqref{eq:explicit_expression_of_rotation_gate}. The Givens rotation
matrix $r_{pq}(\theta_{pq})$ eliminating the $(p,q)$-th entry of a matrix
is of form
\begin{equation*}
        \begin{pmatrix}
            I_{p-1} & & & & \\
            & \cos \theta_{pq} & & \sin\theta_{pq} & \\
            & & I_{q-p-1} & & \\
            & -\sin \theta_{pq} & & \cos\theta_{pq} & \\
            & & & & I_{N-q}
        \end{pmatrix}
\end{equation*}
for $p < q$.
Recall the logarithm of a rotation matrix admits, 
\begin{equation*}
    \log
    \begin{pmatrix}
        \cos \theta & \sin\theta \\
        -\sin \theta & \cos\theta
    \end{pmatrix}
    =
    \begin{pmatrix}
        0 & \theta \\
        -\theta & 0
    \end{pmatrix}.
\end{equation*}
Hence, the logarithm of $r_{pq}(\theta_{pq})$ is
\begin{equation*}
    \log r_{pq}(\theta_{pq}) = \theta_{pq} (E_{pq} - E_{qp}),
\end{equation*}
where $E_{ab}$ denote a zero matrix with a single one at the $(a,b)$-th
position. Substituting the logarithm of $r_{pq}(\theta_{pq})$ into
$U(r_{pq}(\theta_{pq}); \psi)$, we obtain,
\begin{align}
    & U(r_{pq}(\theta_{pq}); \psi) \\
    =\,& \exp({\theta_{pq} \sum_{i,j=1}^n {(E_{pq} - E_{qp})}_{ij}
    \ba^\dag(\psi_i) \ba(\psi_j)}) \\
    =\,& \exp({\theta_{pq}(\ba^\dag(\psi_p) \ba(\psi_q)
    - \ba^\dag(\psi_q) \ba(\psi_p))}),
\end{align}
which proves \eqref{eq:explicit_expression_of_rotation_gate}.

Now \eqref{eq:encoded_phase} and \eqref{eq:encoded_givens_rotation} can be
derived based on \eqref{eq:explicit_expression_of_phase_gate} and
\eqref{eq:explicit_expression_of_rotation_gate}, respectively.

By matrix function property and
\eqref{eq:explicit_expression_of_phase_gate}, the Jordan-Wigner encoded
phase rotation admits,
\begin{equation*}
    \begin{split}
        & J_\psi \bP_p(\phi; \psi) J_\psi^{-1} \\
        =\,& J_\psi \exp(-\imath \phi \bn(\psi_p)) J_\psi^{-1} \\
        =\,& \exp( -\imath \phi J_\psi \bn(\psi_p) J_\psi^{-1}).
    \end{split}
\end{equation*}
The Jordan-Wigner encoding of the number operator could be derived from
that of creation and annihilation operators as in \eqref{eq:JWadagaop},
\begin{equation*}
    \begin{split}
        &J_\psi \bn(\psi_p) J_\psi^{-1}\\
        =& J_\psi \ba^\dag(\psi_p) \ba(\psi_p) J_\psi^{-1} \\
        =& J_\psi \ba^\dag(\psi_p) J_\psi^{-1} J_\psi \ba(\psi_p)
        J_\psi^{-1} \\
        =& I_1 \otimes \cdots\otimes I_{p-1} \otimes (\ket{1} \bra{0})(\ket{0} \bra{1})
        \otimes I_{p+1} \otimes\cdots\otimes I_n \\
        =& I_1 \otimes \cdots\otimes I_{p-1} \otimes
        \begin{pmatrix}
            0 & 0 \\
            0 & 1
        \end{pmatrix}
        \otimes I_{p+1} \otimes\cdots\otimes I_n \\
    \end{split}
\end{equation*}
Hence, we have,
\begin{equation*}
    \begin{split}
        & J_\psi \bP_p(\phi; \psi) J_\psi^{-1} \\
        =\,& \exp( I_1 \otimes \cdots\otimes I_{p-1} \otimes
        \begin{pmatrix}
            0 & 0 \\
            0 & -\imath \phi
        \end{pmatrix}
        \otimes I_{p+1} \otimes\cdots\otimes I_n) \\
        = & I_1 \otimes \cdots\otimes I_{p-1} \otimes 
        \begin{pmatrix}
            1 & 0 \\
            0 & e^{-\imath \phi}
        \end{pmatrix}
        \otimes I_{p+1} \otimes\cdots\otimes I_n.
    \end{split}
\end{equation*}

Similarly, by matrix function property and
\eqref{eq:explicit_expression_of_rotation_gate}, we could derive the
Jordan-Wigner encoded Givens rotation. For simplicity, we give the
derivation for $p = q-1$ in detail, which is the only Givens rotation used
in \cite{kivlichan2018}. We omit the subscript in $\theta_{pq} =
\theta_{q-1,q}$ in the following, and the Jordan-Wigner encoded Givens
rotation admits,
\begin{equation*}
    \begin{split}
        & J_\psi \bR_{q-1,q}(\theta;\psi) J_\psi^{-1} \\
        =\,& \exp(J_\psi \left( \theta (\ba^\dag(\psi_{q-1}) \ba(\psi_q)
        - \ba^\dag(\psi_q) \ba(\psi_{q-1})) \right) J_\psi^{-1}) \\
        = & \exp(I_1 \otimes \cdots\otimes I_{q-2} \otimes
        \begin{pmatrix}
            0 & 0 & 0 & 0 \\
            0 & 0 & \theta & 0 \\
            0 & -\theta & 0 & 0 \\
            0 & 0 & 0 & 0
        \end{pmatrix}
        \otimes\cdots\otimes I_n) \\
        = & I_1 \otimes \cdots\otimes I_{q-2} \otimes 
        \begin{pmatrix}
            1 & 0 & 0 & 0 \\
            0 & \cos\theta & \sin\theta & 0 \\
            0 & -\sin\theta & \cos\theta & 0 \\
            0 & 0 & 0 & 1
        \end{pmatrix}
        \otimes\cdots\otimes I_n.\\
    \end{split}
\end{equation*}

\end{document}